\documentclass[runningheads]{llncs}
\usepackage[T1]{fontenc}
\usepackage{graphicx}
\usepackage{multirow}
\usepackage{booktabs}
\usepackage{bbding}
\usepackage{hyperref}
\usepackage[normalem]{ulem}
\useunder{\uline}{\ul}{}
\usepackage[table,xcdraw]{xcolor}

\hypersetup{
   colorlinks,
   linkcolor={green!50!black},
   citecolor={red!50!black},
   urlcolor={blue!80!black}
}
\begin{document}
\title{Effective Segmentation of Post-Treatment Gliomas Using Simple Approaches: Artificial Sequence Generation and Ensemble Models}
%
%
\author{Heejong Kim\inst{2}\thanks{Equal contribution} \and
Leo Milecki\inst{2}$^*$ \and
Mina C Moghadam\inst{2}$^*$ \and
Fengbei Liu\inst{1}$^*$ \and
Minh Nguyen\inst{1}$^*$ \and
Eric Qiu\inst{1}$^*$ \and
Abhishek Thanki\inst{3}$^*$ \and
Mert R Sabuncu\inst{1,2}}
\authorrunning{H Kim et al.}
%
\institute{School of Electrical and Computer Engineering, Cornell University and Cornell Tech, New York, USA \and
Department of Radiology, Weill Cornell Medicine, New York, USA \and
Weill Cornell Graduate School of Medical Sciences, New York, USA
}

\maketitle              
\begin{abstract}
Segmentation is a crucial task in the medical imaging field and is often an important primary step or even a prerequisite to the analysis of medical volumes. Yet treatments such as surgery complicate the accurate delineation of regions of interest. The BraTS Post-Treatment 2024 Challenge published the first public dataset for post-surgery glioma segmentation and addresses the aforementioned issue by fostering the development of automated segmentation tools for glioma in MRI data. In this effort, we propose two straightforward approaches to enhance the segmentation performances of deep learning-based methodologies. First, we incorporate an additional input based on a simple linear combination of the available MRI sequences input, which highlights enhancing tumors. Second, we employ various ensembling methods to weigh the contribution of a battery of models. Our results demonstrate that these approaches significantly improve segmentation performance compared to baseline models, underscoring the effectiveness of these simple approaches in improving medical image segmentation tasks.

\keywords{BraTS 2024 \and Glioma Post Treatment \and Ensemble \and Segmentation}
\end{abstract}

\section{Introduction}
Gliomas, the most prevalent malignant primary brain tumors in adults, present significant clinical challenges due to their diffuse nature and variability in biological behavior. Among gliomas, diffuse gliomas are particularly problematic because of their infiltrative growth patterns within the central nervous system, which complicates treatment and monitoring~\cite{ostrom2022}~\cite{louis2021}. These tumors often exhibit a range of responses to therapy and have varied prognoses, necessitating a multi-modal approach to treatment that includes surgery, radiation therapy, and systemic therapies. Despite these efforts, effective management and outcome prediction remain challenging.

Magnetic Resonance Imaging (MRI) is the cornerstone of post-treatment imaging for diffuse gliomas. It provides essential insights into tumor size, location, and morphological changes over time, which are crucial for evaluating treatment response and guiding subsequent clinical decisions. Accurate segmentation of gliomas from MRI scans is therefore critical for assessing residual tumor volume, planning further interventions, and predicting patient outcomes.

Several research works have been focusing on tumor segmentation tasks to accurately detect and delineate brain tumors~\cite{menze2014}~\cite{bakas2018}. While significant progress has been made in developing segmentation algorithms for gliomas, most existing works have focused on pre-treatment or general glioma segmentation~\cite{chang2019}~\cite{lotan2022}~\cite{gooya2012}~\cite{hussain2018}. Few works were emphasizing the specific challenges associated with post-treatment imaging~\cite{lotan2022}~\cite{sorensen2023}~\cite{rudie2022}~\cite{tang2020}. This gap underscores the need for specialized approaches that address the unique characteristics of post-treatment MRI scans, where residual tumor and treatment effects can be difficult to distinguish.

The 2024 BraTS challenge focuses on post-treatment gliomas and the development of data-driven models for the semantic segmentation of different tumor regions \cite{de2024}. The challenge's dataset includes post-treatment MRI data for diffuse gliomas, introducing a supplementary sub-region known as the ‘resection cavity,’ left as a result of surgery, a new feature compared to previous BraTS challenges. The main objective of the challenge is to monitor disease progression after surgery and to further help in guiding treatment decisions. Additionally, the dataset and algorithms provided in the challenge can serve as a foundational resource for future research aimed at differentiating treatment modifications from residual or recurrent tumors, forecasting outcomes, and assessing treatment responses.

In this work, we address the task of segmenting post-treatment gliomas using the BraTS 2024 challenge dataset. The challenge consists of effectively delineating tumor regions after surgery, which introduces complexities not present in pre-treatment imaging. We hypothesize that incorporating additional input modalities and applying ensemble techniques will enhance segmentation outcomes. To test this hypothesis, we explore straightforward methods such as generating new sequences like T1Gd-T1 to better highlight different tumor regions and employing ensemble models like STAPLE and weighted averaging of baseline model predictions. Our results reveal that these methods can significantly improve the segmentation of post-treatment gliomas, demonstrating their potential to advance the field of medical image analysis.

\section{Methods}

\subsection{Dataset}
This retrospective study includes approximately 2,200 patients from seven academic clinical centers across the United States. The patients have been diagnosed with diffuse gliomas and have undergone various treatments, including surgery, radiation therapy, and additional therapeutic interventions~\cite{de2024}. The MRI scans provided by the BraTS challenge for these patients are available in NIfTI format and encompass multiple imaging modalities: 1) pre-contrast T1-weighted (T1), 2) post-contrast T1-weighted (T1Gd), 3) T2-weighted (T2), and 4) T2 Fluid Attenuated Inversion Recovery (FLAIR) volumes. The ground truth data was produced through pre-processing steps on expert annotations that included co-registering the images to a standard anatomical template, interpolating to a uniform resolution of $1 mm^3$, and performing skull stripping. The sub-region labels are manually annotated by radiologists. The labels include enhancing tissue (ET), non-enhancing tumor core (NETC), surrounding non-enhancing FLAIR hyperintensity (SNFH), and resection cavity (RC). We used the dataset as provided in the original challenge's training dataset. 

\subsubsection{Additional Input Modality (T1Gd-T1)}
\begin{figure}[hbt!]
    \centering
    \includegraphics[width=0.6\textwidth]{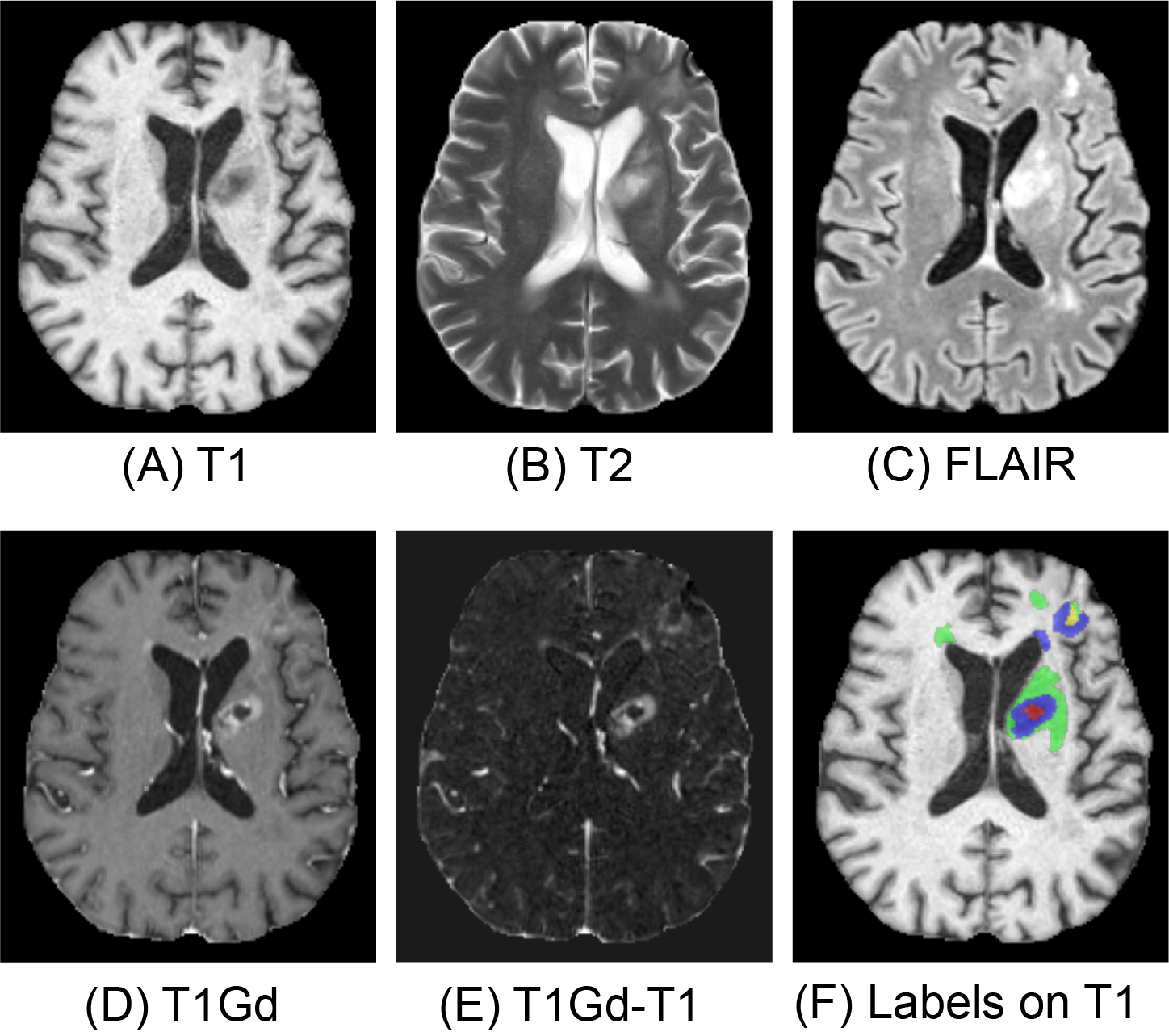} 
    \caption{Four MR imaging modalities from the 2024 BraTS challenge dataset (A-D) and a calculated modality (T1Gd-T1, E). Tumor sub-region labels (F) include enhancing tissue (ET, blue), non-enhancing tumor core (NETC, red), surrounding non-enhancing FLAIR hyperintensity (SNFH, green), and resection cavity (RC, yellow).}
    \label{fig:dataset}
\end{figure}

During the curation of the 2024 challenge dataset, annotators were provided with the T1 contrast subtraction (T1Gd-T1) image for segmentation~\cite{de2024}. Inspired by this, we used the T1Gd - T1 image as an additional input. In Figure~\ref{fig:dataset}, T1Gd-T1 highlights the ET.

\subsection{Architecture}
We based our models on mainly three different network architectures: 1) nnUNet - which is based on the original UNet architecture, 2) nnUNet ResEnc - which is an extended version of nnUNet making use of the residual connections in the encoder, and 3) SegResNet - which is a CNN based encoder-decoder architecture incorporating a variational autoencoder technique~\cite{isensee2021nnu}~\cite{myronenko20193d}. 

\subsubsection{nnUNet:}
nnU-Net is a segmentation framework that configures and trains a U-Net model~\cite{isensee2021nnu}. U-Net consists of an encoder-decoder network where the encoder preserves the semantic information whilst reducing the spatial dimensions and the decoder then reconstructs the segmentation map by upsampling the information obtained from the encoder as well as the corresponding spatial information received through the skip connections \cite{ronneberger2015u}. We trained the baseline nnUnet with the default configuration on the 3D full-resolution data. The network was trained using the nnU-Net framework with the following configuration:  batch size set to 2, patch size set to (128, 160, 112), and the median image size in voxels set to (142, 175, 136). 

\subsubsection{nnUNet ResEnc:}
nnUNet ResEnc in the nnU-Net framework utilizes U-net with residual skip connections in the encoder part of the network~\cite{isensee2021nnu}. We specifically used the newly introduced nnU-Net ResEnc presets which have the ability to adapt the batch and patch sizes depending on the VRAM budget. We trained the L and XL versions of the nnUNet ResEnc with the default configuration on 3D full-resolution data. In the case of L configuration, the batch size was 3, the patch size was (160, 192, 160), and the median image size in voxels was (142, 175, 136). In the case of XL configuration, the batch size was 5, the patch size was (160, 192, 160), and the median image size in voxels was (142, 175, 136). 

\subsubsection{SegResNet:}
SegResNet is based on an encoder-decoder architecture but extends it with an additional variational autoencoder (VAE) part~\cite{myronenko20193d}. The encoder part utilizes ResNet blocks along with group normalization instead of batch normalization~\cite{he2016identity}. The decoder part is similar to the encoder; however, it is based on a single block for each spatial level. The VAE part reduces the encoder output to a low-dimensional space, followed by sampling from a Gaussian distribution. The sampled data is then reconstructed into an image using a decoder-like model, with the key difference being the absence of inter-level skip connections from the encoder. The network was trained using the nnU-Net framework on 3D full-resolution data with the following configuration: batch size set to 2, patch size set to (128, 160, 112), and the median image size in voxels set to (142, 175, 136). 

\subsection{Training}
All four models were trained on two groups of input data: 1) 4 input scans (T1, T1Gd, T2, and T2 FLAIR), and 2) 5 input scans (T1, T1Gd, T2, T2 FLAIR, and T1Gd-T1). 
We utilized an 80/20 split for training and validation from the entire dataset. Following the nnU-Net framework, a variety of data augmentation techniques were applied during training: rotations, scaling, Gaussian noise, Gaussian blur, brightness, contrast, simulation of low resolution, gamma correction, and mirroring. 
All models were trained for 1000 epochs using the SGD optimizer with Nesterov momentum~\cite{Sutskever2013-OnLearning} ($\mu=0.99$) with a starting learning rate of $0.01$ following a polynomial schedule on a single NVIDIA GPU using Pytorch~\cite{Paszke2019-Pytorch}.

\subsection{Test time augmentation}
We utilized test-time augmentation to enhance the robustness of predictions by averaging results from various augmented versions of the input data. This approach helps to account for potential variations and uncertainties in unseen data. In this work, we employed the default nnUNet test-time augmentations \cite{isensee2021nnu} during inference. It includes mirroring (flipping) along different axes and applying Gaussian weighting of the predictions, helping to smooth the output and reduce boundary artifacts.

\subsection{Ensemble}
The ensembling of predictions can yield a significant boost in prediction performance and has been employed successfully by winners of previous editions of the BraTS challenges~\cite{zeineldin2022multimodal,ferreira2024we}.
We explored two different ways of ensembling model predictions: STAPLE~\cite{rohlfing2004performance,warfield2004simultaneous} and weighted average. 
STAPLE~\cite{rohlfing2004performance,warfield2004simultaneous} constructs a weighted average of the predictions but does so without requiring held-out data. Instead, STAPLE estimates the weights assigned to each model using the model predictions themselves and the EM algorithm~\cite{dempster1977maximum}.
When held-out data are available, one could also output a weighted probabilities average where the weights are estimated by maximizing the ensemble performance on the held-out data. One advantage of this approach is that we can weight models for multi-labels separately.

\begin{table}[ht]
\centering
\scalebox{0.81}{
\begin{tabular}{l|c|c|c|cccccc}
\hline
\multicolumn{1}{c|}{$\uparrow$} & \begin{tabular}[c]{@{}c@{}}Additional\\ channel input\end{tabular} & Ensemble                       & Method               & ET                                   & NETC                                   & RC                                   & SNFH                                 & TC                                   & WT                                   \\ \hline
\#1                             &                                                                    &                                & nnUNet               & 0.7717                               & 0.8386                                 & 0.7697                               & 0.8365                               & 0.7747                               & 0.8330                               \\
\#2                             &                                                                    &                                & SegResNet            & 0.7887                               & 0.8265                                 & 0.7799                               & 0.8252                               & 0.7856                               & 0.8254                               \\
\#3                             &                                                                    &                                & nnUNet + ResEncUNetL & 0.8047                               & 0.8248                                 & 0.7777                               & 0.8273                               & 0.8027                               & 0.8289                               \\
\#4                             & \multirow{-4}{*}{\XSolidBrush}                                     &                                & nnUNet + ResEncUNetXL & 0.8032                               & 0.8431                                 & \textbf{\color[HTML]{036400}0.8019}                               & 0.8393                               & 0.8035                               & 0.8392                               \\ \cline{1-2} \cline{4-10} 
\#5                             &                                                                    &                                & nnUNet               & 0.7985                               & 0.8426                                 & 0.7779                               & 0.8276                               & 0.7979                               & 0.8295                               \\
\#6                             &                                                                    &                                & SegResNet            & 0.7869                               & 0.8303                                 & 0.7798                               & 0.8199                               & 0.7900                               & 0.8179                               \\
\#7                             &                                                                    &                                & nnUNet + ResEncUNetL & 0.8038                               & 0.8435                                 & 0.7784                               & 0.8244                               & 0.8024                               & 0.8238                               \\
\#8                             & \multirow{-4}{*}{\CheckmarkBold}                                   & \multirow{-8}{*}{\XSolidBrush} & nnUNet + ResEncUNetXL & \textbf{\color[HTML]{9A0000} 0.8136} & 0.8409                                 & 0.7781                               & 0.8331                               & \textbf{\color[HTML]{9A0000} 0.8161} & 0.8349                               \\ \hline
\#9                             & \XSolidBrush                                                       &                                & Ensemble \#1-4       & \textbf{\color[HTML]{036400} 0.8106}                               & 0.8398                                 & \textbf{\color[HTML]{9A0000} 0.8076} & 0.8348                               & 0.8051                               & 0.8320                               \\
\#10                            & \CheckmarkBold                                                     &                                & Ensemble \#5-9       & 0.8095                               & {\color[HTML]{9A0000} \textbf{0.8512}} & 0.7917                               & 0.8314                               & \textbf{\color[HTML]{036400}0.8128}                               & 0.8316                               \\
\#11                            & \CheckmarkBold                                                     & \multirow{-3}{*}{STAPLE}       & Ensemble \#1-9       & 0.8036                               & 0.8466                                 & 0.7999                               & 0.8377                               & 0.8090                               & 0.8341                               \\ \hline
\#12                            & \XSolidBrush                                                       &                                & Ensemble \#1-4       & 0.8100                               & 0.8423                                 & \textbf{\color[HTML]{9A0000} 0.8076} & \textbf{\color[HTML]{036400}0.8422}                               & 0.8077                               & \textbf{\color[HTML]{9A0000} 0.8436} \\
\#13                            & \CheckmarkBold                                                     &                                & Ensemble \#5-9       & 0.8104                               & 0.8427                                 & 0.8016                               & 0.8421                               & 0.8081                               & \textbf{\color[HTML]{036400}0.8435}                               \\
\#14                            & \CheckmarkBold                                                     & \multirow{-3}{*}{Weighted}     & Ensemble \#1-9       & 0.8071                               & \textbf{\color[HTML]{036400}0.8470}                                 & 0.8024                               & \textbf{\color[HTML]{9A0000} 0.8427} & 0.8093                               & 0.8420                               \\ \hline
\end{tabular}}
\caption{LD scores for internal validation set with different baselines. Additional channel input indicates the usage of T1Gd-T1.  \textcolor[HTML]{9A0000}{\textbf{Red}} is the best performing result and \textcolor[HTML]{036400}{\textbf{green}} is the second best. Result is the higher the better.}
\label{tab:internal_val_dice}
\end{table}

\begin{table}[h]
\centering
\scalebox{0.73}{
\begin{tabular}{l|c|c|c|llllll}
\hline
\multicolumn{1}{c|}{$\downarrow$} & \begin{tabular}[c]{@{}c@{}}Additional\\ channel input\end{tabular} & Ensemble                       & Method               & \multicolumn{1}{c}{ET}                  & \multicolumn{1}{c}{NETC}                & \multicolumn{1}{c}{RC}                  & \multicolumn{1}{c}{SNFH}                & \multicolumn{1}{c}{TC}                  & \multicolumn{1}{c}{WT}                  \\ \hline
\#1                               &                                                                    &                                & nnUNet               & 51.1324                                 & 27.1954                                 & 41.0131                                 & 37.1673                                 & 50.1382                                 & 41.2318                                 \\
\#2                               &                                                                    &                                & SegResNet            & 47.3470                                 & 28.4743                                 & 33.7970                                 & 37.7961                                 & 47.4018                                 & 39.2164                                 \\
\#3                               &                                                                    &                                & nnUNet + ResEncUNetL & {\color[HTML]{036400}\textbf{40.3632} }                                & 30.7466                                 & 37.7362                                 & 35.1327                                 & 41.8607                                 & 37.1744                                 \\
\#4                               & \multirow{-4}{*}{\XSolidBrush}                                     &                                & nnUNet + ResEncUNetXL & 46.2630                                 & 23.3088                                 & 33.5624                                 & {\color[HTML]{036400}\textbf{30.6463 }}                                & 45.7200                                 & 34.0185                                 \\ \cline{1-2} \cline{4-10} 
\#5                               &                                                                    &                                & nnUNet               & 42.6100                                 & 23.4603                                 & 37.8478                                 & 37.1878                                 & 43.9434                                 & 37.1065                                 \\
\#6                               &                                                                    &                                & SegResNet            & 44.4186                                 & 24.7710                                 & 33.1014                                 & 41.7299                                 & 46.1129                                 & 46.0196                                 \\
\#7                               &                                                                    &                                & nnUNet + ResEncUNetL & 41.3596                                 & 23.2999                                 & 35.8579                                 & 36.2353                                 & 44.3718                                 & 39.7798                                 \\
\#8                               & \multirow{-4}{*}{\CheckmarkBold}                                   & \multirow{-8}{*}{\XSolidBrush} & nnUNet + ResEncUNetXL & 40.4495                                 & {\color[HTML]{036400}\textbf{22.4659} }                                & 34.6625                                 & 37.7991                                 & {\color[HTML]{9A0000} \textbf{40.9828}} & 38.4813                                 \\ \hline
\#9                               & \XSolidBrush                                                       &                                & Ensemble \#1-4       & 41.1699                                 & 27.5377                                 & 28.8558                                 & 34.8404                                 & 43.9045                                 & 38.1087                                 \\
\#10                              & \CheckmarkBold                                                     &                                & Ensemble \#5-9       & 41.8326                                 & {\color[HTML]{9A0000} \textbf{20.7410}} & 32.1034                                 & 36.5352                                 & 41.5172                                 & 36.8008                                 \\
\#11                              & \CheckmarkBold                                                     & \multirow{-3}{*}{STAPLE}       & Ensemble \#1-9       & 40.9159                                 & 22.8680                                 & 31.0782                                 & {\color[HTML]{9A0000} \textbf{30.6314}} & 42.0808                                 & {\color[HTML]{9A0000} \textbf{33.6783}} \\ \hline
\#12                              & \XSolidBrush                                                       &                                & Ensemble \#1-4       & 40.6045                                 & 25.7191                                 & {\color[HTML]{9A0000} \textbf{29.4378}} & 35.2625                                 & 41.2713                                 & 37.1799                                 \\
\#13                              & \CheckmarkBold                                                     &                                & Ensemble \#5-9       & {\color[HTML]{9A0000} \textbf{39.4369}} & 25.7088                                 & 31.9253                                 & 35.2629                                 & {\color[HTML]{036400}\textbf{41.2662}}                                 & 37.1792                                 \\
\#14                              & \CheckmarkBold                                                     & \multirow{-3}{*}{Weighted}     & Ensemble \#1-9       & 42.3202                                 & 23.2703                                 & {\color[HTML]{036400}\textbf{29.6432 }}                                & 31.9311                                 & 42.8091                                 & {\color[HTML]{036400}\textbf{33.9768   }}                              \\ \hline
\end{tabular}}
\caption{LH95 scores for internal validation set with different baselines. Additional channel input indicates the usage of T1Gd-T1.  \textcolor[HTML]{9A0000}{\textbf{Red}} is the best performing result and \textcolor[HTML]{036400}{\textbf{green}} is the second best. The metric is the lower, the better.}
\label{tab:internal_val_hau95}
\end{table}

\begin{table}[]
\centering
\begin{tabular}{l|c|c|cccccc}
\hline
\multicolumn{1}{c|}{$\uparrow$}   & Ensemble                   & Method         & ET                                     & NETC                                   & RC                                     & SNFH                                   & TC                                     & WT                                     \\ \hline
\#15 &                            & Ensemble \#1-4 & 0.7292                                 & 0.7855                                 & {\color[HTML]{9A0000} \textbf{0.7014}} & 0.8478                                 & {\color[HTML]{9A0000} \textbf{0.7200}} & 0.8500                                 \\
\#16 &                            & Ensemble \#5-9 & 0.7193                                 & 0.7836                                 & 0.6843                                 & 0.8486                                 & 0.7008                                 & 0.8465                                 \\
\#17 & \multirow{-3}{*}{STAPLE}   & Ensemble \#1-9 & 0.7277                                 & {\color[HTML]{9A0000} \textbf{0.7868}} & 0.6994                                 & 0.8454                                 & 0.7154                                 & 0.8446                                 \\ \hline
\#18 &                            & Ensemble \#1-4 & {\color[HTML]{036400}\textbf{0.7332}}                                 & {\color[HTML]{036400}\textbf{0.7866}}                                 & {\color[HTML]{036400}\textbf{0.7009}}                                 & {\color[HTML]{036400}\textbf{0.8594}}                                 & {\color[HTML]{036400}\textbf{0.7173}}                                 & {\color[HTML]{036400}\textbf{0.8601}}                                 \\
\#19 &                            & Ensemble \#5-9 & {\color[HTML]{9A0000} \textbf{0.7334}} & 0.7824                                 & 0.6956                                 & {\color[HTML]{036400}\textbf{0.8594}}                                 & 0.7166                                 & 0.8600                                 \\
\#20 & \multirow{-3}{*}{Weighted} & Ensemble \#1-9 & {\color[HTML]{036400}\textbf{0.7332}}                                 & 0.7861                                 & 0.6948                                 & {\color[HTML]{9A0000} \textbf{0.8682}} & 0.7120                                 & {\color[HTML]{9A0000} \textbf{0.8704}} \\ \hline
\end{tabular}
\caption{LD scores for hold-out validation set with ensemble methods. \textcolor[HTML]{9A0000}{\textbf{Red}} is the best performing result and \textcolor[HTML]{036400}{\textbf{green}} is the second best. }
\label{tab:online_val}
\end{table}

\begin{table}[]
\centering
\begin{tabular}{l|c|c|cccccc}
\hline
\multicolumn{1}{c|}{$\downarrow$}& Ensemble                   & Method         & \multicolumn{1}{c}{ET}                  & \multicolumn{1}{c}{NETC}                & \multicolumn{1}{c}{RC}                  & \multicolumn{1}{c}{SNFH}                & \multicolumn{1}{c}{TC}                  & \multicolumn{1}{c}{WT}                  \\ \hline
\#15       &                            & Ensemble \#1-4 & 43.3998                                 & 44.4383                                 & {\color[HTML]{036400} \textbf{54.3403}} & 31.4488                                 & {\color[HTML]{9A0000} \textbf{42.9108}} & 32.3994                                 \\
\#16       &                            & Ensemble \#5-9 & 52.9701                                 & 45.9666                                 & 60.3439                                 & 31.3575                                 & 56.8107                                 & 33.1450                                 \\
\#17       & \multirow{-3}{*}{STAPLE}   & Ensemble \#1-9 & 46.0162                                 & {\color[HTML]{036400} \textbf{45.6417}} & {\color[HTML]{9A0000} \textbf{53.7277}} & 33.7174                                 & 51.5435                                 & 34.9169                                 \\ \hline
\#18       &                            & Ensemble \#1-4 & {\color[HTML]{036400}\textbf{42.6223}}                                 & {\color[HTML]{9A0000} \textbf{44.2384}} & 54.7714                                 & {\color[HTML]{036400} \textbf{28.7516}}                                 & 46.7207                                 & 29.9510                                 \\
\#19       &                            & Ensemble \#5-9 & {\color[HTML]{9A0000} \textbf{42.6185}} & 44.2957                                 & 56.5904                                 & 28.7519                                 & {\color[HTML]{036400} \textbf{46.6246}}                                 & {\color[HTML]{036400} \textbf{29.9491}}                                 \\
\#20       & \multirow{-3}{*}{Weighted} & Ensemble \#1-9 & 44.4066                                 & 46.0825                                 & 56.1886                                 & {\color[HTML]{9A0000} \textbf{25.4669}} & 50.3338                                 & {\color[HTML]{9A0000} \textbf{25.6896}} \\ \hline
\end{tabular}
\caption{LH95 scores for hold-out validation set with ensemble methods. \textcolor[HTML]{9A0000}{\textbf{Red}} is the best performing result and \textcolor[HTML]{036400}{\textbf{green}} is the second best.}
\label{tab:online_val_hau95}
\end{table}

\section{Results}

\begin{figure}[]
    \centering
    \includegraphics[width=\textwidth]{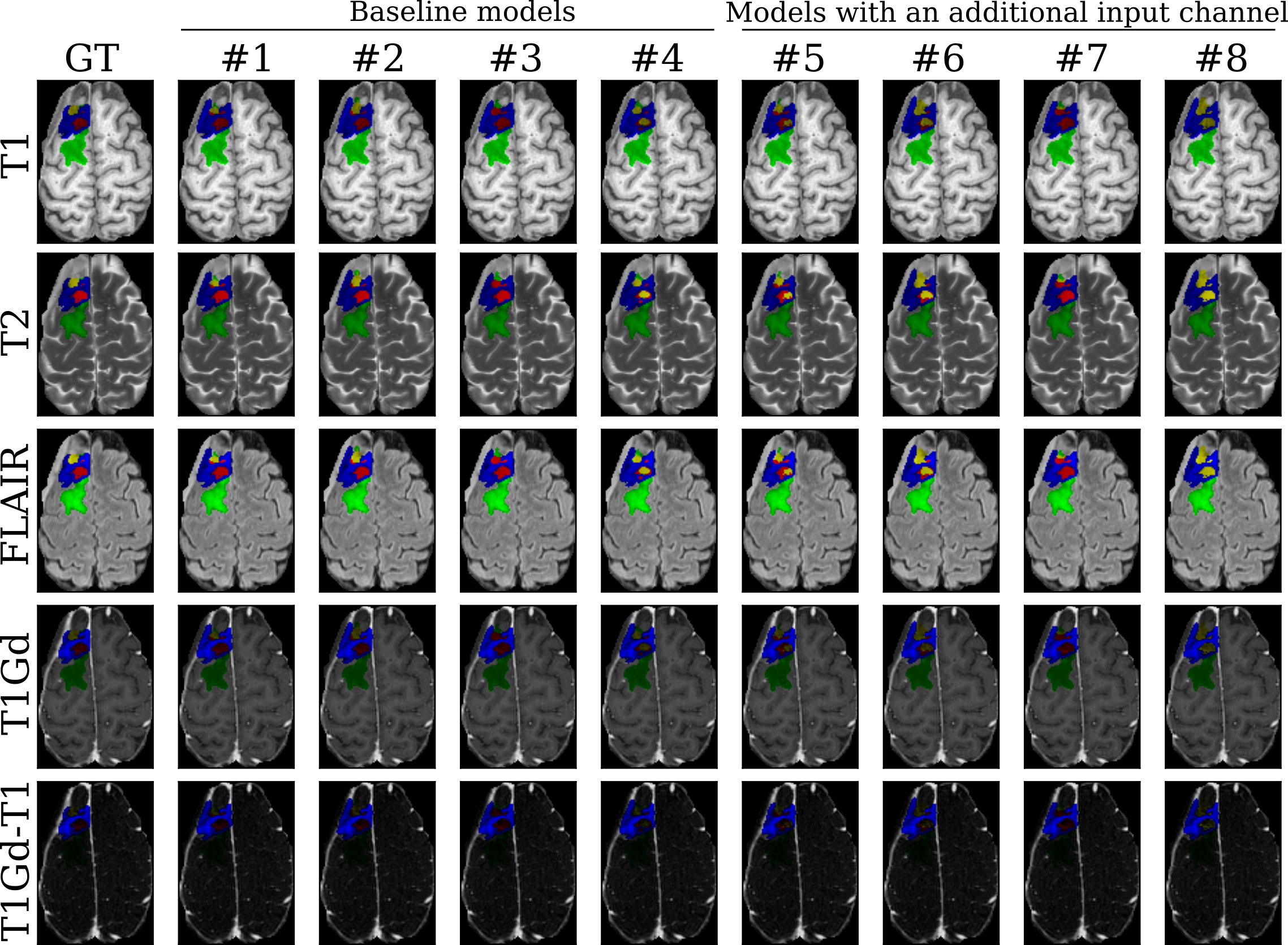} 
    \caption{Segmentation results visualization on one subject in the internal validation set on the different MR imaging input modalities (lines). The ground truth annotation (GT) is compared against our baseline models (columns \#1 to \#4) and our models with the T1Gd-T1 input (columns \#5 to \#8). Labels include enhancing tissue (ET, blue), non-enhancing tumor core (NETC, red), surrounding non-enhancing FLAIR hyperintensity (SNFH, green), and resection cavity (RC, yellow).}
    \label{fig:result-internal-single-model}
\end{figure}

\begin{figure}[]
    \centering
    \includegraphics[width=0.8\textwidth]{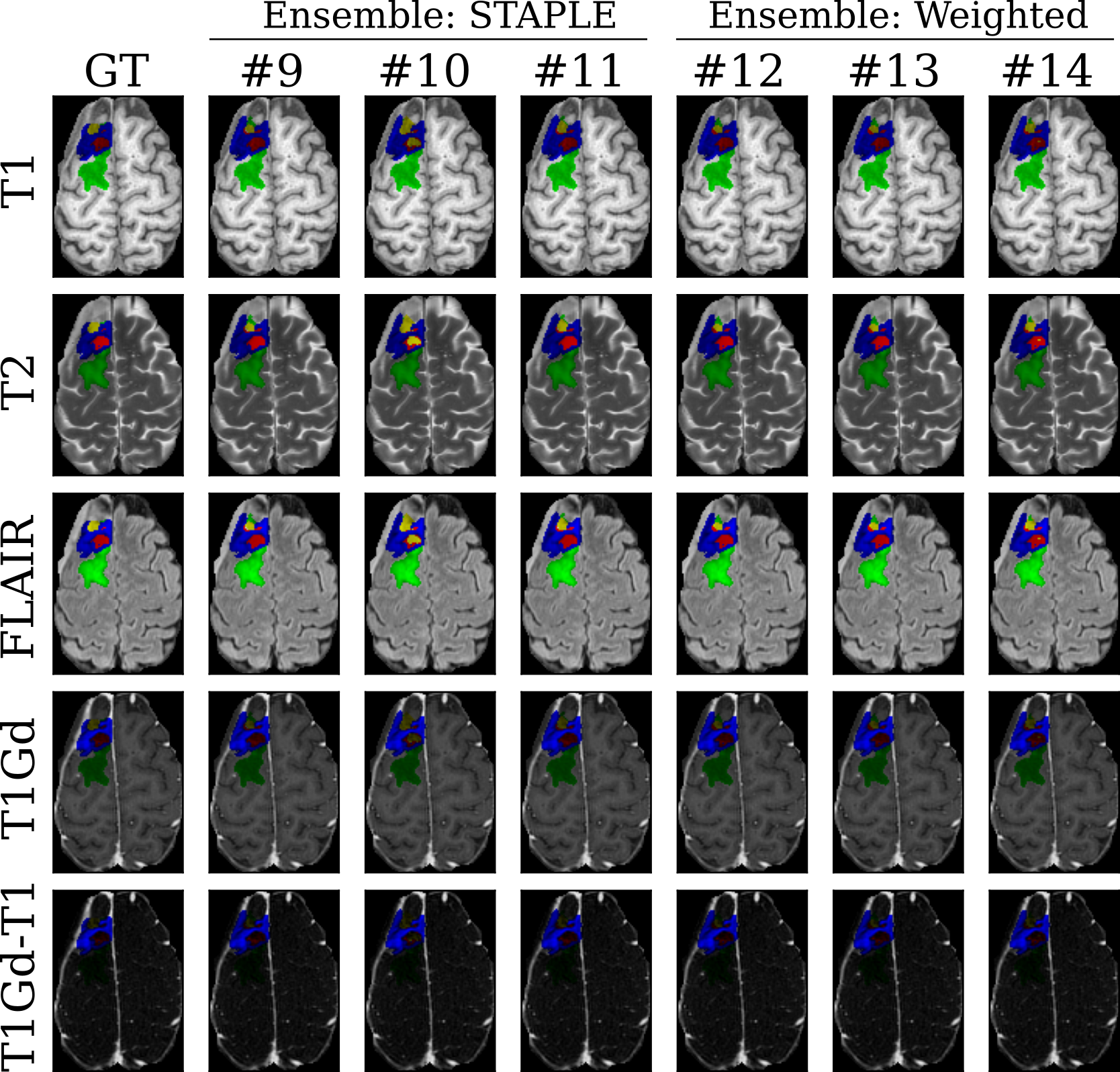} 
    \caption{Segmentation results visualization on one subject in the internal validation set on the different MR imaging input modalities (lines). The ground truth annotation (GT) is compared against our ensemble models using STAPLE (columns \#9 to \#11) and the proposed weighted approach (columns \#12 to \#14). Labels include enhancing tissue (ET, blue), non-enhancing tumor core (NETC, red), surrounding non-enhancing FLAIR hyperintensity (SNFH, green), and resection cavity (RC, yellow).}
    \label{fig:result-internal-ensemble-model}
\end{figure}

In this section, we show our results on the internal validation set ($N=280$) and hold-out validation set ($N=188$) with different baselines and ensembling approaches. We follow the provided metrics\footnote{\url{https://github.com/rachitsaluja/BraTS-2024-Metrics.git}} and calculate Lesion-wise Dice scores (LD) and Lesion-wise Hausdorff95 scores (LH95). For each label, the Dice and Hausdorff distance at the 95\% percentile metrics are computed separately on unique lesions. The different lesions are detected using morphological operators and connected components analysis.

For the internal validation set, we show results for LD and LH95 scores in Table~\ref{tab:internal_val_dice} and Table~\ref{tab:internal_val_hau95}. For index \#1-4, we calculate LD and LH95 for each baseline with 4 input scans. For index \#5-9, we calculate LD and LH95 with 5 input scans, including the additional T1Gd-T1 input scan. 
Figure~\ref{fig:result-internal-single-model} presents qualitative segmentation results for one subject in the internal validation set for the baseline models with (columns \#1-4) and without (columns \#5-8) including the additional T1Gd-T1 input scan against the ground truth annotations (GT). While both groups of models depict similar predictions for the SNFH label (green), we observe more accurate contours for the ET (blue) using the proposed additional input. 
For indices \#9-11 and \#12-14, we calculate STAPLE/weighted ensemble results with respect to the aforementioned baselines. We observe that for single baseline results, incorporating additional channel input generally improves LD scores. We also observe better LD and LH95 scores with larger baseline models (ResEncUNetL to ResEncUNetXL). For ensemble approaches, STAPLE and weighted average generally improve performance compared with single baselines. 
In Figure~\ref{fig:result-internal-ensemble-model}, we visualize qualitative segmentation results for one subject in the internal validation set for ensemble models using STABLE (columns \#9-11) and the proposed weighted approach (columns \#12-14) against the ground truth annotations (GT). For the selected subject, the STABLE ensemble models provide a better detection of the RC (yellow) while the proposed weighted ensemble models focus on accurate segmentation of the NETC (red). 

For the hold-out validation set, we show results in Table~\ref{tab:online_val} and Table~\ref{tab:online_val_hau95}. We submit only the ensemble approaches as they generally outperform single models in the internal validation set. On the hold-out validation set, we observe that the weighted average generally performs better than STAPLE on LD scores. We also observe that without T1Gd-T1 (Ensemble \#1-4) and with T1Gd-T1 (Ensemble \#5-9) perform similarly in both LD and LH95. However, ensembling them (Ensemble \#1-9) provides significant improvement in RC and WT classes, with 1\% LD improvement and 3-4\% LH95 reduction. This validates our argument that additional input modalities and ensemble techniques can lead to improved segmentation outcomes.

\section{Discussion}
In this work, we proposed a simple yet effective approach for the BraTS 2024 Segmentation Challenge - Adult Glioma Post Treatment. 
We demonstrate that simple approaches, such as artificially generating new sequences like T1Gd-T1 to enhance different tumor regions and utilizing ensemble models like STAPLE and weighted averaging, can effectively improve segmentation performance in the post-treatment glioma dataset. 

Several papers have previously reported the usefulness of subtraction images for segmentation tasks~\cite{duan2008segmentation,leung1998value}, as well as integrating multiple MRI modalities for diverse medical imaging-based tasks~\cite{Calhoun2016-MultimodalIllness}. However, there is a lack of studies on using subtraction images for deep learning models. Our analysis of the synthesized T1Gd-T1 input image demonstrated the potential utility of subtraction images for glioma sub-region segmentation tasks. 

Aggregating predictions from multiple models can significantly boost performance, as well as demonstrate its relevance for more robust and generalizable capabilities~\cite{Yang2021-OnTC}. In addition to STAPLE, a long-established method, we demonstrated that a label-wise weighted averaging technique can outperform STAPLE. 

These findings suggest that integrating diverse data sources and leveraging ensemble techniques can significantly improve the accuracy and reliability of glioma segmentation models. Future work will focus on exploring additional input channels and ensembling strategies, as well as investigating the potential of these methods in other segmentation tasks and medical imaging applications.



%
%
%
%

\bibliographystyle{splncs04}
\bibliography{bib}

\end{document}